\documentclass{article}
\usepackage{dcase2018,cite}
\usepackage{graphicx,amsmath,amssymb,multirow,dblfloatfix,siunitx, bbold}

\title{A multi-room reverberant dataset for \\ sound event localization and detection}

\name{Sharath Adavanne, Archontis Politis, and Tuomas Virtanen\thanks{This work has received funding from the European Research Council under the ERC Grant Agreement 637422 EVERYSOUND.}}
\address{Audio Research Group, Tampere University, Finland}

\begin{document}
\ninept
\maketitle

\begin{abstract}
This paper presents the sound event localization and detection (SELD) task setup for the DCASE 2019 challenge. The goal of the SELD task is to detect the temporal activities of a known set of sound event classes, and further localize them in space when active. As part of the challenge, a synthesized dataset with each sound event associated with a spatial coordinate represented using azimuth and elevation angles is provided. These sound events are spatialized using real-life impulse responses collected at multiple spatial coordinates in five different rooms with varying dimensions and material properties. A baseline SELD method employing a convolutional recurrent neural network is used to generate benchmark scores for this reverberant dataset. The benchmark scores are obtained using the recommended cross-validation setup.
\end{abstract}


\vspace{-5pt}
\section{Introduction}\vspace{-5pt}
The goals of the sound event localization and detection (SELD) task includes recognizing a known set of sound event classes such as `dog bark', `bird call', and `human speech' in the acoustic scene, detecting their individual onset and offset times, and further localizing them in space when active. Such a SELD method can automatically describe human and social activities with a spatial dimension, and help machines to interact with the world more seamlessly. Specifically, SELD can be an important module in assisted listening systems, scene information visualization systems, immersive interactive media, and spatial machine cognition for scene-based deployment of services.

The number of existing methods for the SELD task are limited~\cite{Butko2011,Chakraborty_ICASSP2014,Hirvonen2015,Lopatka2016,Adavanne_JSTSP2018} providing ample research opportunity. Thus to promote the research and development of SELD methods we propose to organize the SELD task at DCASE 2019\footnote{http://dcase.community/challenge2019/task-sound-event-localization-and-detection}. Previously, the SELD task has been treated as two stand-alone tasks of sound event detection (SED) and direction of arrival (DOA) estimation~\cite{Butko2011}. The SED in~\cite{Butko2011} was performed using a classifier-based on Gaussian mixture model - hidden Markov model, and the DOA estimation using the steered response power (SRP). In the presence of multiple overlapping sound events, this approach resulted in the data association problem of assigning the individual sound events detected in a time-frame to their corresponding DOA locations. This data association problem was overcome in~\cite{Chakraborty_ICASSP2014} by using a sound-model-based localization instead of the SRP method. 

Recent methods have proposed to jointly learn the SELD sub-tasks of SED and DOA estimation using deep neural networks (DNN). Based on the DOA estimation approach, these methods can be broadly categorized into classification~\cite{Hirvonen2015} and regression approaches~\cite{Adavanne_JSTSP2018}. The classification approaches estimate a discrete set of angles, whereas the regression approaches estimate continuous angles. As the classification approach, Hirvonen~\cite{Hirvonen2015} employed a convolutional neural network and treated SELD as a multiclass-multilabel classification task. Power spectrograms extracted from multichannel audio were used as the acoustic feature and mapped to two sound classes at eight different azimuth angles. Formally, the SELD task was performed by learning an acoustic model, parameterized by parameters $\mathbf{W}$, that estimates the probability of each sound class to be active at a certain time-frame, and discrete spatial angle $P(\mathbf{Y}| \mathbf{X}, \mathbf{W})$, where $\mathbf{X} \in \mathbb{R}^{K\times T \times F}$ is the frame-wise acoustic feature for each of the $K$ channels of audio with feature-length $F$, and number of time-frames $T$. $\mathbf{Y} \in \mathbb{R}^{T\times C \times U}$ is the class-wise SELD probabilities for $C$ sound classes and $U$ number of azimuth angles. The SELD activity could then be obtained from the class-wise probabilities $\mathbf{Y}$ by applying a binary threshold. Finally, the onset and offset times of the individual sound event classes, and their respective azimuth locations could be obtained from the presence of prediction in consecutive time-frames.


As the regression approach, we recently proposed a convolutional recurrent neural network, SELDnet~\cite{Adavanne_JSTSP2018}, that was shown to perform significantly better than~\cite{Hirvonen2015}. In terms of acoustic features $\mathbf{X}$, the SELDnet employed the naive phase and magnitude components of the spectrogram, thereby avoiding any task- or method-specific feature extraction. These features were mapped to two outputs using a joint acoustic model $\mathbf{W}$. As the first output, SED was performed as a multiclass-multilabel classification by estimating the class-wise probabilities $\mathbf{Y}_{\text{SED}} \in \mathbb{R}^{T\times C}$ as $P(\mathbf{Y}_{\text{SED}} | \mathbf{X}, \mathbf{W})$. The second output, DOA estimation was performed as multioutput regression task by estimating directly the $L$ dimensional spatial location as $\mathbf{Y}_{\text{DOA}} \in \mathbb{R}^{T\times L\times G\times C}$ for each of the $C$ classes as $f_{\mathbf{W}}: \mathbf{X} \mapsto \mathbf{Y}_{\text{DOA}}$. At each time-frame, $G$ spatial coordinates are estimated per sound class and can be chosen based on the complexity of the sound scene and recording array setup capabilities. In~\cite{Adavanne_JSTSP2018}, one trajectory was estimated per sound class ($G=1$), and the respective DOA was represented using its 3D Cartesian coordinates along $x$, $y$, and $z$ axes ($L = 3$).

The two DNN-based approaches for SELD, i.e., classification~\cite{Hirvonen2015} and regression approach~\cite{Adavanne_JSTSP2018}, have their respective advantages and restrictions. For instance, the resolution of DOA estimation in a classification approach is limited to the fixed set of angles used during training, and the performance on unseen DOA values is unknown. For datasets with a higher number of sound classes and DOA angles, the number of output nodes of the classifier increases rapidly. Training such a large multilabel classifier, where the training labels per frame have a few positives classes representing active sound class and location in comparison to a large number of negative classes, poses problems of imbalanced dataset training. Additionally, such a large output classifier requires a larger dataset to have sufficient examples for each class. On the other hand, the regression approach performs seamlessly on unseen DOA values, does not face the imbalanced dataset problems, and can learn from smaller datasets. As discussed earlier, algorithmically the two approaches can potentially recognize multiple instances of the same sound class occurring simultaneously, but such a scenario has never been evaluated and hence their performances are unknown. Additionally, it was observed that the DOA estimation of the classification approach for seen locations was more accurate than the regression approach. This was concluded to be a result of incomplete learning of the regression mapping function due to the small size of dataset~\cite{Adavanne_JSTSP2018}. 

Data-driven SELD approaches~\cite{Hirvonen2015,Adavanne_JSTSP2018} require sufficient data with annotation of sound event activities and their spatial location. Annotating such a real-life recording to produce large dataset is a tedious task. One of the ways to overcome this is to develop methods that can learn to perform real-life SELD from a large synthesized dataset and a smaller real-life dataset. The performance of such methods is directly related to the similarity of the synthesized dataset to the real-life sound scene. In~\cite{Adavanne_JSTSP2018} we proposed to create such realistic sound scene by convolving real-life impulse responses with real-life isolated sound events, and summing them with real-life ambient sound. These sound scenes were created to have both isolated, and overlapping sound events. The ambient sound was added to the recording at different signal-to-noise ratios (SNRs) to simulate varying real-life conditions. However, all the impulse responses for the dataset in~\cite{Adavanne_JSTSP2018} were collected from a single environment. 
Learning real-life SELD with such restricted dataset is difficult. One of the approaches to overcome this is to train the methods with larger acoustic variability in the training data. In this regard, for the DCASE 2019 SELD task, we employ impulse responses collected from five different environments with varying room dimensions and reverberant properties. Additionally, in order to support research focused on specific audio formats, we provide an identical dataset in two formats of four-channels each: first-order Ambisonics and microphone array recordings.

To summarize, we propose the SELD task for the DCASE 2019 challenge to promote SELD research. We present a challenging multi-room reverberant dataset\footnote{https://doi.org/10.5281/zenodo.2580091} with varying numbers of overlapping sound events, and a fixed evaluation setup to compare the performance of different methods. As the benchmark, we provide a modified version of the recently proposed SELDnet\footnote{https://github.com/sharathadavanne/seld-dcase2019}~\cite{Adavanne_JSTSP2018} and report the results on the multi-room reverberant dataset.

\vspace{-5pt}
\section{Multi-room reverberant Dataset}\vspace{-5pt}

The SELD task in DCASE 2019 provides two datasets, TAU Spatial Sound Events 2019 - Ambisonic (FOA) and TAU Spatial Sound Events 2019 - Microphone Array (MIC), of identical sound scenes with the only difference in the format of the audio. The FOA dataset provides four-channel First-Order Ambisonic recordings while the MIC dataset provides four-channel directional microphone recordings from a tetrahedral array configuration. Both formats are extracted from the same microphone array. The SELD methods can be developed on either one of the two or both the datasets to exploit their mutual information. Both the datasets, consists of a development and evaluation set. The development set consists of 400 one-minute long recordings sampled at 48000 Hz, divided into four cross-validation splits of 100 recordings each. The evaluation set consists of 100 one-minute long recordings. These recordings were synthesized using spatial room impulse response (IRs) collected from five indoor environments, at 504 unique combinations of azimuth-elevation-distance. In order to synthesize these recordings the collected IRs were convolved with isolated sound events from DCASE 2016 task 2\footnote{http://dcase.community/challenge2016/task-sound-event-detection-in-synthetic-audio}. Additionally, half the number of recordings have up to two temporally overlapping sound events, and the remaining have no overlapping. Finally, to create a realistic sound scene recording, natural ambient noise collected in the IR recording environments was added to the synthesized recordings such that the average SNR of the sound events was 30 dB. The only explicit difference between each of the development dataset splits and evaluation dataset is the isolated sound event examples employed.

\vspace{-5pt}
\subsection{Real-life Impulse Response Collection}\vspace{-5pt}
The real-life IR recordings were collected using an Eigenmike\footnote{https://mhacoustics.com/products\#eigenmike} spherical microphone array. A Genelec G Two loudspeaker\footnote{https://www.genelec.com/home-speakers/g-series-active-speakers} was used to playback a maximum length sequence (MLS) around the Eigenmike. The MLS playback level was ensured to be 30 dB greater than the ambient sound level during the recording. The IRs were obtained in the STFT domain using a least-squares regression between the known measurement signal (MLS) and far-field recording independently at each frequency. These IRs were collected in the following directions: a) 36 IRs at every $\ang{10}$ azimuth angle, for 9 elevations from $\ang{-40}$ to $\ang{40}$ at $\ang{10}$ increments, at 1 m distance from the Eigenmike, resulting in 324 discrete DOAs. b) 36 IRs at every $\ang{10}$ azimuth angle, for 5 elevations from $\ang{-20}$ to $\ang{20}$ at $\ang{10}$ increments, at 2 m distance from the Eigenmike, resulting in 180 discrete DOAs. The IRs were recorded at five different indoor environments inside the Tampere University campus at Hervanta, Finland during non-office hours. These environments had varying room dimensions, furniture, flooring and roof materials. Additionally, we also collected 30 minutes of ambient noise recordings from these five environments with the IR recording setup unchanged during office hours thereby obtaining realistic ambient noise. We refer the readers to the DCASE 2019 challenge webpage for description on individual environments.

\vspace{-5pt}
\subsection{Dataset Synthesis}\vspace{-5pt}
The isolated sound events dataset from DCASE 2016 task 2 consists of 11 classes, each with 20 examples. These examples were randomly split into five sets with an equal number of examples for each class; the first four sets were used to synthesize the four splits of the development dataset, while the remaining one set was used for the evaluation dataset. Each of the one-minute recordings were generated by convolving randomly chosen sound event examples with a corresponding random IR to spatially position them at a given distance, azimuth and elevation angles. The IRs chosen for each recording are all from the same environment. Further, these spatialized sound events were temporally positioned using randomly chosen start times following the maximum number of overlapping sound events criterion. Finally, ambient noise collected at the respective IR environment was added to the synthesized recording such that the average SNR of the sound events is 30 dB.

Since the number of channels in the IRs is equal to the number of microphones in Eigenmike (32), in order to create the MIC dataset we select four microphones that have a nearly-uniform tetrahedral coverage of the sphere. Those are the channels 6, 10, 26, and 22  that corresponds to microphone positions ($\ang{45}$, $\ang{35}$, 4.2 cm), ($\ang{-45}$, $\ang{-35}$, 4.2 cm), ($\ang{135}$, $\ang{-35}$, 4.2 cm) and ($\ang{-135}$, $\ang{35}$, 4.2 cm). The spherical coordinate system in use is right-handed with the front at ($\ang{0}$, $\ang{0}$), left at ($\ang{90}$, $\ang{0}$) and top at ($\ang{0}$, $\ang{90}$). Finally, the FOA dataset is obtained by converting the 32-channel microphone signals to the first-order Ambisonics format, by means of encoding filters based on anechoic measurements of the Eigenmike array response, generated with the methods detailed in \cite{politis2017comparing}.

\vspace{-5pt}
\subsection{Array Response}\vspace{-5pt}
For model-based localization approaches the array response may be considered known. The following theoretical spatial responses (steering vectors) modeling the two formats describe the directional response of each channel to a source incident from DOA given by azimuth angle $\phi$ and elevation angle $\theta$.

For the FOA format, the array response is given by the real orthonormalized spherical harmonics:
\begin{eqnarray}
H_1(\phi, \theta, f) &=& 1 \\
H_2(\phi, \theta, f) &=& \sqrt{3} * \sin(\phi) * \cos(\theta) \\
H_3(\phi, \theta, f) &=& \sqrt{3} * \sin(\theta) \\
H_4(\phi, \theta, f) &=& \sqrt{3} * \cos(\phi) * \cos(\theta).
\end{eqnarray}

For the tetrahedral array of microphones mounted on spherical baffle, similar to Eigenmike, an analytical expression for the directional array response is given by the expansion:
\begin{eqnarray}
H_m(\phi_m, \theta_m, \phi, \theta, \omega) = \nonumber\\ \frac{1}{(\omega R/c)^2}\sum_{n=0}^{30} \frac{i^{n-1}}{h_n'^{(2)}(\omega R/c)}(2n+1)P_n(\cos(\gamma_m)),
\end{eqnarray}
where $m$ is the channel number, $(\phi_m, \theta_m)$ are the specific microphone's azimuth and elevation position, $\omega = 2\pi f$ is the angular frequency, $R = 0.042$ m is the array radius, $c = 343$ m/s is the speed of sound, $\cos(\gamma_m)$ is the cosine angle between the microphone position and the DOA, $P_n$ is the unnormalized Legendre polynomial of degree $n$, and $h_n'^{(2)}$ is the derivative with respect to the argument of a spherical Hankel function of the second kind. The expansion is limited to 30 terms which provide negligible modeling error up to 20 kHz. Note that the Ambisonics format is frequency-independent, something that holds true to about 9 kHz for Eigenmike and deviates gradually from the ideal response for higher frequencies.

\begin{figure}[t]
  \centering
  \centerline{\includegraphics[height=9.5cm,keepaspectratio, trim=0.1cm 1.1cm 0.1cm 0.25cm,clip]{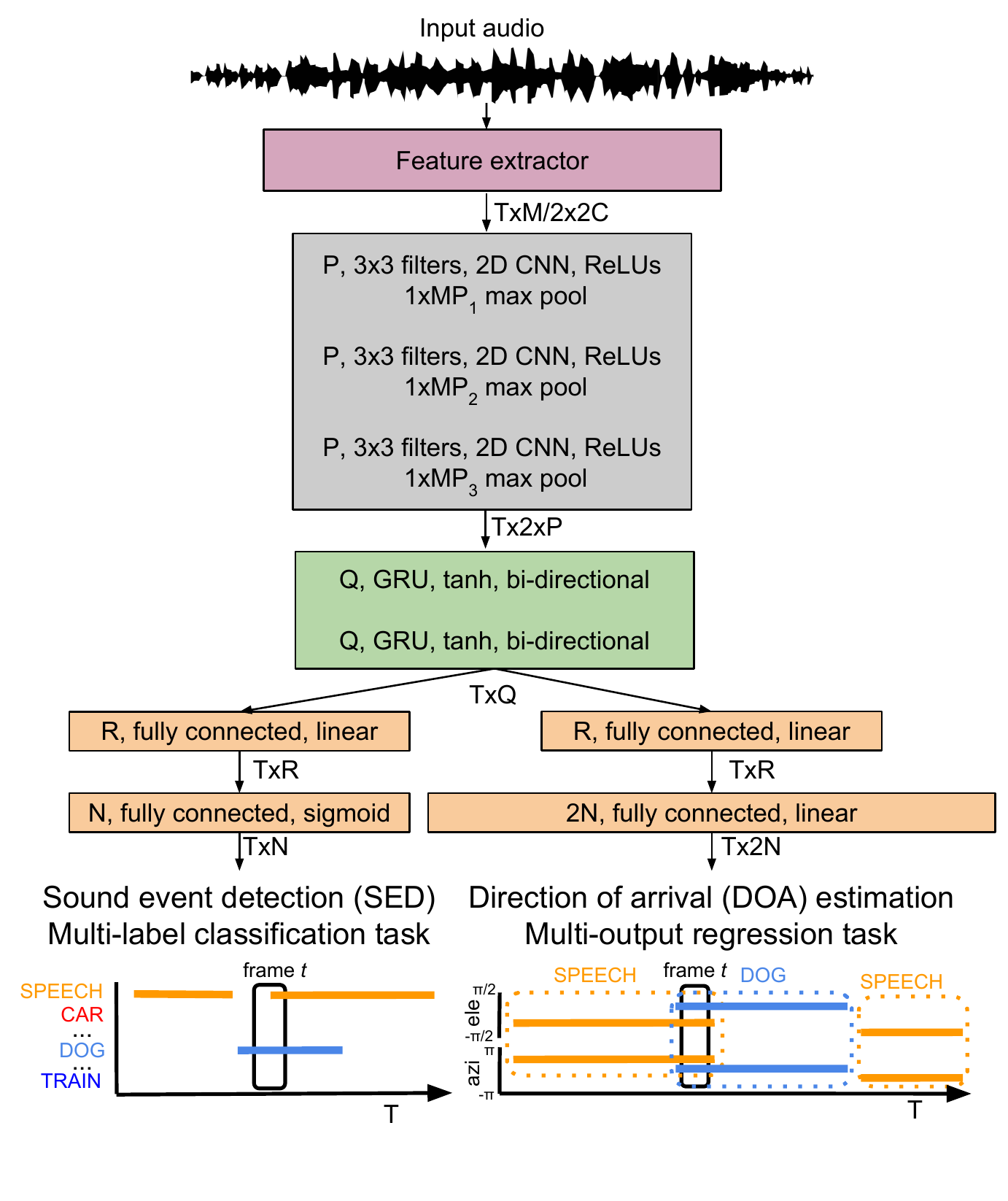}} 
  \vspace{-5pt}
  \caption{Convolutional recurrent neural network for SELD.}
  \label{fig:crnn}
  \vspace{-15pt}
\end{figure}

\vspace{-5pt}
\section{Baseline Method}\vspace{-5pt}
As the benchmark method, we employ the SELDnet~\cite{Adavanne_JSTSP2018}. Contrary to~\cite{Adavanne_JSTSP2018}, where the DOA estimation is performed as multi-output regression of the 3D Cartesian DOA vector components $x, y, z \in [-1 , 1]$, in this benchmark implementation we directly estimate the azimuth $\phi \in [-\pi, \pi]$ and elevation $\theta \in [-\pi/2, \pi/2]$ angles. Accordingly, the activation function of the DOA estimation layer is changed from tanh to linear. The remaining architecture remains unchanged and is illustrated in Figure~\ref{fig:crnn}. The input to the method is a multichannel audio of 48 kHz sampling rate, from which the phase and magnitude components of the spectrogram are extracted using 2048-point discrete Fourier transform from 40 ms length Hanning window and 20 ms hop length. A sequence of $T$ spectrogram frames ($T$ = 128) is then fed to the three convolutional layers that extract shift-invariant features using 64 filters each. Batch normalization is used after each convolutional layer. Dimensionality reduction of the input spectrogram feature is performed using max pooling operation only along the frequency axis. The temporal axis is untouched to keep the resolution of the output unchanged from the input dimension.

\begin{table*}[!b]
\centering  
\renewcommand\thetable{3}
\vspace{-15pt}
\caption{SELDnet performance on overlapping sound events and reverberant scenes. ID - identifier, DE - DOA Error, FR - Frame Recall}
\resizebox{\textwidth}{!}{
\begin{tabular}{l|l|cccc|cccc|cccc|cccc}
\multicolumn{2}{c}{} & \multicolumn{8}{c|}{Development dataset scores} & \multicolumn{8}{c}{Evaluation dataset scores} \\ \cline{3-18}
\multicolumn{2}{c|}{} & \multicolumn{4}{c|}{FOA} & \multicolumn{4}{c|}{MIC} & \multicolumn{4}{c|}{FOA} & \multicolumn{4}{c}{MIC} \\ \cline{3-18}
\multicolumn{1}{c}{} & ID & ER & F & DE & FR & ER & F & DE & FR & ER & F & DE & FR & ER & F & DE & FR \\ \cline{2-18}
\multirow{2}{*}{Overlap} & 1 & \textbf{0.32} & \textbf{82.2} & \textbf{23.1} & \textbf{93.0} & \textbf{0.35} & \textbf{81.2} & \textbf{25.9} & \textbf{92.3} & &&&&&&& \\
 & 2 & 0.35 & 78.6 & 31.6 & 77.8 &  \textbf{0.35} & 79.1 & 33.6 & 75.8 & &&&&&&&\\ \hline
\multirow{5}{*}{\begin{tabular}[c]{@{}l@{}}Impulse\\response\end{tabular}} & 1 & \textbf{0.30} & \textbf{81.6} & \textbf{28.3} & 85.3  & \textbf{0.33} & 80.2 & 30.4 & 83.9  & &&&&&&& \\
 & 2 & 0.38 & 78.6 & 28.5 & \textbf{86.7} &0.36 & 80.2 & 30.7 & \textbf{86.2} & &&&&&&&\\
 & 3 & 0.33 & 80.0 & 28.7 & 84.3  & 0.35 & 79.9 & 30.7 & 82.5 &&&&&&&&\\
 & 4 & 0.37 & 79.3 & \textbf{28.3} & 84.9& 0.35 & \textbf{80.4 }& \textbf{30.3} & 83.7 &&&&&&&& \\
 & 5 & 0.34 & 80.0 & 29.0 & 85.7& 0.36 & 79.5 & 31.9 & 83.4 & &&&&&&&\\ \hline
\multicolumn{2}{l|}{Total} & 0.34 & 79.9 & 28.5 & 85.4 & 0.35 & 80.0 & 30.8 & 84.0 & &&&&&&&
\end{tabular}
} 
\label{T:ir_ov_results}
\end{table*}

The temporal structure of the sound events is modeled using two bi-directional recurrent layers with 128 gated recurrent units each. Finally, the output of the recurrent layer is shared between two fully-connected layer branches each producing the SED as multiclass multilabel classification and DOA as multi-output regression; together producing the SELD output. The SED output obtained is the class-wise probabilities for the $C$ classes in the dataset at each of the $T$ frames of input spectrogram sequence, resulting in a dimension of $T\times C$. The localization output estimates one single DOA for each of the $C$ classes at every time-frame $T$, i.e., if multiple instances of the same sound class occur in a time frame the SELDnet localizes either one or oscillates between multiple instances. The overall dimension of localization output is $T\times2C$, where $2C$ represents the class-wise azimuth and elevation angles. A sound event class is said to be active if its probability in SED output is greater than the threshold of 0.5, otherwise, the sound class is considered to be absent. The presence of sound class in consecutive time-frames gives the onset and offset times, and the corresponding DOA estimates from the localization output gives the spatial location with respect to time.

A cross-entropy loss is employed for detection output, while a mean square error loss on the spherical distance between reference and estimated locations is employed for the localization output. The combined convolutional recurrent neural network architecture is trained using Adam optimizer and a weighted combination of the two output losses. Specifically, the localization output is weighted $\times50$ more than the detection output.

\begin{table}[!t]
\centering  
\renewcommand\thetable{1}
\vspace{-5pt}
\caption{Cross-validation setup}
\begin{tabular}{l|ccc}
 & \multicolumn{3}{c}{Splits} \\ \cline{2-4}
Folds & Training & Validation & Testing \\ \hline
Fold 1 & 3, 4 & 2 & 1 \\
Fold 2 & 4, 1 & 3 & 2 \\
Fold 3 & 1, 2 & 4 & 3 \\
Fold 4 & 2, 3 & 1 & 4
\end{tabular}
\label{T:splits}\vspace{-15pt}
\end{table}

\vspace{-5pt}
\section{Evaluation}\vspace{-5pt}
\subsection{Evaluation Setup}\vspace{-5pt}

The development dataset consists of four cross-validation splits as shown in Table~\ref{T:splits}. Participants are required to report the performance of their method on the testing splits of the four folds. The performance metrics are calculated by accumulating the required statistics from all the folds~\cite{Forman2010}, and not as the average of the metrics of the individual folds. For the evaluation dataset, participants are allowed to decide the training procedure, i.e. the amount of training and validation files in the development dataset and the number of ensemble models.

\vspace{-5pt}
\subsection{Metrics}\vspace{-5pt}
The SELD task is evaluated with individual metrics for SED and DOA estimation. For SED, we use the F-score and error rate (ER) calculated in one-second segments~\cite{metrics}. For DOA estimation we use two frame-wise metrics~\cite{Adavanne2018_EUSIPCO}: DOA error and frame recall. The DOA error is the average angular error in degrees between the predicted and reference DOAs. For a recording of length $T$ time frames, let $\mathbf{DOA}^t_R$ be the list of all reference DOAs at time-frame $t$ and $\mathbf{DOA}^t_E$ be the list of all estimated DOAs. The DOA error is now defined as
\begin{equation}
DOA\,error = \frac{1}{\sum_{t=1}^{T}{D^t_E}}\sum_{t=1}^{T}{\mathcal{H}(\mathbf{DOA}^t_R, \mathbf{DOA}^t_E)},
\end{equation}
where $D^t_E$ is the number of DOAs in $\mathbf{DOA}^t_E$ at $t$-th frame, and $\mathcal{H}$ is the Hungarian algorithm for solving assignment problem, i.e., matching the individual estimated DOAs with the respective reference DOAs. The Hungarian algorithm solves this by estimating the pair-wise costs between individual predicted and reference DOA using the spherical distance between them, $\sigma = \arccos(\sin\phi_{E}\sin\phi_{R} + \cos\phi_{E}\cos\phi_{R}\cos(\lambda_{R}-\lambda_{E}))$, where the reference DOA is represented by the azimuth angle $\phi_R \in [-\pi, \pi)$ and elevation angle $\lambda_R \in [-\pi/2, \pi/2]$, and the estimated DOA is represented with $(\phi_E, \lambda_E)$ in the similar range as reference DOA.

In order to account for time frames where the number of estimated and reference DOAs are unequal, we report a frame recall type metric, which is calculated as, $Frame\,recall =\sum_{t=1}^{T}{\mathbb{1}(D^t_R = D^t_E)}/T$, with $D^t_R$ the number of DOAs in $\mathbf{DOA}^t_R$ at $t$-th frame, $\mathbb{1}()$ the indicator function returning one if the $(D^t_R = D^t_E)$ condition is met and zero otherwise. The submitted methods will be ranked individually for all four metrics of SED and DOA estimation, and the final positions will be obtained using the cumulative minimum of the ranks.

\begin{table}[!t]
\centering  
\vspace{-5pt}
\renewcommand\thetable{2}
\caption{Evaluation scores for cross-validation folds.}
\label{T:split_results}
\resizebox{\columnwidth}{!}{
\begin{tabular}{c|llll|llll}
\multicolumn{1}{c}{} & \multicolumn{4}{c|}{FOA} & \multicolumn{4}{c}{MIC} \\ \cline{2-9}
Fold & ER & F & DE & FR & ER & F & DE & FR \\ \hline
1 & \textbf{0.25} & \textbf{85.0} & 30.4 & \textbf{86.6} & 0.32 & 81.9 & 32.0 & 84.5  \\
2 & 0.39 & 77.0 & \textbf{25.9} & 85.0 & 0.37 & 79.0 & 30.5 & 83.1 \\
3 & 0.30 & 83.1 & 27.9 & \textbf{86.6} & \textbf{0.29} & \textbf{83.4} & 30.9 & \textbf{85.4}  \\
4 & 0.42 & 74.7 & 30.2 & 83.3 &  0.42 & 76.2 & \textbf{29.8} & 83.1 
\end{tabular}
}\vspace{-15pt}
\end{table}

\vspace{-5pt}
\section{Results}\vspace{-5pt}

The results obtained with the SELDnet for different folds of the development dataset are presented in Table~\ref{T:split_results}. Although the folds are identical for the FOA and MIC datasets, the SELDnet is observed to perform better on fold 1 for FOA and fold 3 for MIC datasets. This suggests that the spectral and spatial information in the two formats are not identical and potentially methods can benefit from mutual information from the two datasets.

The overall results with respect to different numbers of overlapping sound events and different reverberant environments are presented in Table~\ref{T:ir_ov_results}. The general performance of SELDnet on FOA dataset is marginally better than MIC dataset. The SELDnet is seen to perform better when there is no polyphony across datasets. Finally, the SELDnet trained with five environments is seen to perform the best in the first environment.

\vspace{-5pt}
\section{Conclusion}\vspace{-5pt}
In this paper, we proposed the sound event localization and detection (SELD) task for the DCASE 2019 challenge to promote SELD research. An acoustically challenging multi-room reverberant dataset is provided for the task. This dataset is synthesized with isolated sound events that are spatially positioned using real-life impulse responses collected from five-different rooms that have different acoustic properties. Additionally, in order to support research focused on specific audio-formats, the dataset provides four-channel Ambisonic and microphone array recordings of identical sound scenes. Further, the dataset provides a pre-defined four-fold cross-validation split for evaluating the performance of competing methods. As the baseline results for the dataset, we report the performance of a benchmark SELD method based on convolutional recurrent neural network. 

\bibliographystyle{IEEEtran}
\bibliography{template}
\end{document}